\centerline{\bf Optical Flashes Preceding GRBs}
\vskip 0.5cm
\centerline{Bohdan Paczy\'nski}
\centerline{E-mail: bp@astro.princeton.edu}
\centerline{Princeton University Observatory, 124 Peyton Hall, Princeton, 
NJ 08544-1001, USA}
\vskip 0.5cm
\centerline{ABSTRACT}
\vskip 0.3cm

Only one optical flash associated with a gamma-ray burst has been detected so
far by ROTSE.  There are also upper limits obtained by several groups for
several bursts.  Recent model calculations indicate a possibility that optical
flash may precede the main GRB.  Such flashes are undetectable in the currently 
popular observing mode, with optical instruments responding to GRB triggers.
There is a need to develop all sky optical monitoring system capable of 
recognizing flashes in real time, and more powerful instruments that could
respond robotically to optical triggers and carry out follow up observations.

\vskip 0.2cm
\centerline{ - - - - - - - - }
\vskip 0.2cm

Over the last decade several teams in several countries, with a total of
perhaps $ \sim 100 $ participants, were developing optical instrumentation
capable of a rapid response to GRBs detected by space borne instruments.
So far there is only one detection, the only positive result of several man
- centuries of effort: the truly spectacular optical flash detected by the 
ROTSE team (Akerlof et al. 1999) while the GRB 990123 was still active.  To 
be more precise: the data were recorded robotically in real time, following
the Bacodine (Barthelmy et al. 1998) signal sent by BATSE from space.  However,
the BATSE GRB position had a large error, and ROTSE had to cover almost 1000
square degrees in the sky to make sure the source was included.  This large
area implied a huge number of sources and it appears that ROTSE did not have
software to search for a possible optical transient among them.  Fortunately,
the position was refined first by BeppoSAX (Piro et al. 1999), and next by
optical detection of the afterglow (Odewahn et al. 1999), making it possible
for the ROTSE to identify optical flash which reached 9th magnitude at its
peak, while the source was at cosmological redshift: $ {\rm z = 1.6 } $
(Kelson et al. 1999).

In addition to the one detected optical flash several groups reported upper
limits to possible flashes associated with a dozen or so gamma-ray bursts 
(e.g. Krimm et al. 1996, Park et al. 1999, Kehoe et al. 2001).  There are 
many more groups with similar goals and results, but as far as I know none
of them has real time alert system, capable of identifying an unusual optical
transient within minutes or seconds of its appearance.  There are many optical
discoveries done in near real time and reported within a day, or even within 
hours of the detection.  These include comets, supernovae, microlensing events,
some cataclysmic variables.  The reports of unusual, even spectacular flares 
from ordinary stars remain unverified, as none of them was recognized in real
time (Schaefer 1989, Schaefer et al. 2000).

The possibility that a more or less standard GRB model could generate a low
energy precursor to the proper GRB was considered by Meszaros et al. (2001)
and Beloborodov (2001), though the possibility of generating a strong optical
flash prior to the main GRB was not emphasized.  I was struck by this 
possibility upon reading Beloborodov's paper.  I verified physical 
plausibility of the scenario with Pawan Kumar and Alin Panaitescu.  As a 
result of our discussion a serious attempt has been made to explore the 
optical precursor possibility in considerable quantitative details, and the 
results will be reported soon (Kumar \& Panaitescu, 2001).  

It turns out that there is a broad range of model parameters for which a 
strong optical flash is expected, though it is not likely that current models
can make quantitative predictions derived from the first principles.  It is
the other way around: the detection of optical precursors, or firm upper 
limits to their presence, would put strong constraints on the range of 
parameters for GRB models.  In other words: a search for optical flashes 
independent of GRB triggers would provide an important diagnostics for the 
GRBs and their environments.  The idea is very simple.  If there is enough gas
around the source of gamma-rays some photons are back scattered, interact
with the outward streaming photons, and create an electron -- positron plasma.
This contributes to even more back scattering, and it may lead to a runaway,
with a huge increase of the electron scattering optical depth.  Until the
$ {\rm e^+ - e^- } $ cloud is cleared very few gamma-rays escape, and lower
energy photons dominate the fireball emission.

Some GRBs are known to be very X-ray rich (e.g. Feroci et al. 2001).  It is
conceivable that explosions which give rise to some GRBs can also give rise
to mostly X-ray or even mostly optical events, depending on the environment.
There may be other types of explosions, not related to anything we know so far,
and only new observations may discover them.  In addition to gamma-ray and 
X-ray searches from space, there is a prospect of monitoring TeV sky with 
Milagro (http://www.lanl.gov/milagro/) and the radio sky with Lofar 
(http://www.astron.nl/lofar/).  However, the lowest cost hardware can be 
developed in the optical domain (Nemiroff \& Rafert 2000, Paczy\'nski 2000,
2001, Pojma\'nski 2000).  The main bottleneck appears to be software.

The purpose of this note is to point out that there are theoretical reasons
to expect strong optical flashes to precede the main part of some gamma-ray
bursts (Meszaros et al. 2001, Beloborodov 2001, Kumar \& Panaitescu 2001).  
Only somewhat more speculative are strong optical flashes with gamma-ray 
emission strongly suppressed.  The only way to learn about such phenomena is
by developing a wide angle, or all sky optical monitoring system with multi
-- pixel CCD detectors, $ \sim 1 $ minute time resolution, real time data 
processing with instant recognition of a rapid variability, the on line
verification of the event with a larger robotic instrument, preferably with 
at least some spectral capability.  Several groups or individuals are 
developing such system, or at least they are trying (Pojma\'nski 2001, 
Vestrand 2001).  In subsequent steps the time resolution may be reduced to
$ \sim 1 $ second, and later to below 1 second. 

While optical flashes related to powerful explosions at large redshift may or
may not be readily detected, the enigmatic flares from ordinary stars may be 
verified (Schafer 1989, Schafer et al. 2000), and enormous diversity of 
ordinary and not so ordinary variable stars will be discovered and monitored 
(Paczy\'nski 2000, 2001).  We should not underestimate the scientific 
importance of a long term monitoring of such mundane objects like stars,
quasars, and asteroids, as well as a generation of large, well defined samples
of variables of all kinds.  The most challenging aspect of the task will be 
the development of a robust software.

\vskip 0.2cm

This paper is posted on astro-ph only, and it will not be submitted to any
paper journal.  The author welcomes all critical comments by the readers, 
in particular about missing references.  It is a pleasure to acknowledge 
stimulating discussions with Pawan Kumar and Alin Panaitescu.  This
project was not supported by any grant.

\vskip 0.5cm

\centerline{\bf REFERENCES}

\vskip 0.3cm

Akerloff, C. et al. (ROTSE) 1999, astro-ph/9903271 (Nature, 398, 400)

Barthelmy, S. et al. 1998, http://gcn.gsfc.nasa.gov/gcn/

Beloborodov, A. M. 2001, astro-ph/0103321

Feroci, M. et al. 2001, astro-ph/0108414

Kehoe, R. et al. (ROTSE) 2001, astro-ph/0104208

Kelson, D. D. et al. 1999, IAU Circular No. 7096

Krimm, H. A., Vanderspeck, R. K., \& Ricker, G. R. 1996, A\&AS, 120, 251

Kumar, P., \& Panaitescu, A. 2001, in preparation

Meszaros, P., Ramirez-Ruiz, E., \& Rees, M.J., 2001, astro-ph/0011284 
   (ApJ 554, 660)

Nemiroff, R. J. \& Rafert, J. B. 1999, astro-ph/9809403 (PASP, 111, 886)

Odewahn, S. C. et al. 1999, http://gcn.gsfc.nasa.gov/gcn/gcn3/201.gcn3

Paczy\'nski, B. 2000, astro-ph/0005284 (PASP, 112, 1281)

Paczy\'nski, B. 2001, astro-ph/0108112 (IAU Coll. 183, in press)

Park, H. et al. (LOTIS) 1999, A\&AS, 138, 577

Piro, L. et al. 1999, http://gcn.gsfc.nasa.gov/gcn/gcn3/199.gcn3

Pojma\'nski, G. 2000, astro-ph/0005236 (AcA, 50, 177)

Pojma\'nski, G. 2001, private communication

Schaefer, B. E. 1989, ApJ, 337, 927

Schaefer, B. E., et al. 2000, astro-ph/9909188 (ApJ, 337, 927)

Vestrand, T. 2001, private communication

\vfill \end \bye